# Ensemble Algorithm for the Selection Problem by NMR Ensemble Quantum Computers

Chien-Yuan Chen, Chih-Cheng Hsueh
*Department of Information Engineering, I-Shou University,*
*Kaohsiung Country, Taiwan, 840, R.O.C*

## Abstract

In this paper, we present an ensemble algorithm for selection problem to find the k-th smallest element in the unsorted database. We will search the k-th smallest element by using "divide-and-conquer" strategy. We first divide D, the domain of the database, into two parts, determine which of the two parts the object element sought belongs to, and then concentrate on that part. We repeat divide that part until object element is found. The determination of which part depends on the output of ensemble counting scheme, which outputs the number of assignments satisfying the value of the oracle query function is set to one. Our algorithm modifies the ensemble counting scheme by constructing a new oracle query function $g_y(\cdot)$. We set $g_y(j)$ to one if the j-th element is less than or equal to y. At first, we set y to the middle value of D and perform the ensemble counting scheme with the oracle query function $g_y(\cdot)$ to compute the number C, the number of j satisfying $g_y(j)=1$. If $C>k$, the object element lies in the first half of D. If $C \leq k$, then it must be in the second half of D. We recursively apply this method by adapting y until the object element is found. Our algorithm thus requires $O(\ln|D|)$ oracle queries for adequate measure accuracy to find the k-th smallest element, where $|D|$ denotes the size of D.

## 1. Introduction

NMR (Nuclear Magnetic Resonance) ensemble computing has many applications in ensemble quantum information processing [2, 5-7, 11]. In NMR, there are many identical molecules. Each molecule, like a quantum computer, contains massive different spins, like qubits. Many ensemble algorithms [1, 9, 10, 17] in NMR have been presented. In 2000, Brüschweiler presented an ensemble searching algorithm to discover the object element from the unsorted N elements by $O(\ln N)$ oracle queries. Brüschweiler's algorithm provided an exponential speedup over Grover's quantum search algorithm [16]. It can reduce $O(\sqrt{N})$ to $O(\ln N)$. In 2002, Yang et al. simplified Brüschweiler's algorithm [10] such that an ancillary bit is not needed. Obviously, these algorithms are designed to search known elements. But, some problems, like the media problem, the selection problem, and the maximum problem, need searching unknown elements from an unsorted database.

In this paper, we want to solve the selection problem such that k-th smallest element of an unsorted database can be found. We use the ensemble counting scheme in NMR ensemble computing [2, 5, 7, 13-15] to design our ensemble searching algorithm for finding the k-th smallest element. We use the concept of divide-and-conquer to construct our algorithm. The ensemble counting scheme can output the number of assignments satisfying the value of the oracle query function is set to one. Assume that we want to search the k-th element among N elements in an unsorted database with the integer domain D. Without loss of generality, let $N = 2^n$ for some integer n. We further assume that $|D|$, the size of D, is finite. It means that the number of different elements in D is finite. We prepare two registers: the first register with n qubits and the second register with an ancillary qubit. We then construct an oracle query function $g_y(\cdot)$ such that

$$\begin{cases} g_y(j) = 1, \text{ if the j-th element is less than or equal to y,} \\ g_y(j) = 0, \text{ if the j-th element is larger than y.} \end{cases}$$

The action of $g_y(\cdot)$ can be viewed as a permutation, which will be discussed in Section 4. We, therefore, can construct a unitary operation $U_{g_y}(\cdot)$ corresponding to $g_y(\cdot)$. We apply $g_y(\cdot)$ to the first register and put the result on the second register. At first, we set y to the middle value of D and perform the ensemble counting scheme with the oracle query function $g_y(\cdot)$ to compute the number C, the number of j satisfying $g_y(j) = 1$. If $C > k$, the object element lies in the first half of D. If $C \leq k$, then it must be in the second half of D. We recursively apply this method by adapting y until the object element is found. Our algorithm thus requires $O(\ln|D|)$ oracle queries to find the k-th smallest element.

The remainder of this paper is organized as follows. In Section 2, we review the ensemble counting scheme in NMR ensemble quantum computers. We describe an ensemble searching algorithm for finding the k-th smallest number in NMR ensemble quantum computers in Section 3. In Section 4, we discuss our algorithm. Section 5 draws the conclusions.

## 2. Review of ensemble counting scheme

We give an oracle query function $f:\{0,1\}^n \rightarrow \{0,1\}$ to find C which is the number of assignments satisfying f=1. Let $U_f(\cdot)$ be the unitary operation of the oracle query function $f(\cdot)$. The following ensemble counting scheme [3] can output C by performing the oracle query function $f(\cdot)$ once.

**Ensemble counting scheme: [3]**

Step 1:

First, we prepare two quantum registers: the first register with n qubits and the second register with one ancillary qubit. These two registers are initially set to $|0\rangle$. We apply the Walsh-Hadamard Transformation H to $|0\rangle|0\rangle$. Then, we have

$$H(|0\rangle|0\rangle) \longrightarrow \frac{1}{\sqrt{2^n}} \sum_{j=0}^{2^n-1} |j\rangle|0\rangle.$$

Step 2:

We perform the operation $U_f$ which applies $f(\cdot)$ to the first register and puts the result in the second register. Thus, we get

$$U_f(\frac{1}{\sqrt{2^n}}\sum_{j=0}^{2^n-1}|j\rangle|0\rangle) \to \frac{1}{\sqrt{2^n}}\sum_{j=0}^{2^n-1}|j\rangle|f(j)\rangle.$$

Step 3:

We measure the second register and obtain the output $\alpha$. According to the computational model of a bulk quantum Turing machine [12], $\alpha$ and C satisfy the inequality

$$(\frac{C}{2^n}-\frac{2^n-C}{2^n})-\frac{1}{2^{\varepsilon-1}}<\alpha<(\frac{C}{2^n}-\frac{2^n-C}{2^n})+\frac{1}{2^{\varepsilon-1}},$$

where measurement accuracy $\varepsilon$ is a positive integer. Thus, we can find that C lies between $2^{n-1}(1+\alpha)-2^{n-\varepsilon}$ and $2^{n-1}(1+\alpha)+2^{n-\varepsilon}$. Here, we assume that $\varepsilon \approx n$. We thus output $C=2^{n-1}(1+\alpha)$.

## 3. Our ensemble algorithm for solving the selection problem

Assume that we want to search the k-th smallest element among $N=2^n$ elements, say $a_0, a_1, ..., a_{N-1}$, in an unsorted database with the integer domain D. We further assume that $|D|$ is finite. Let min and max denote the minimum and maximum value in D. First, we define an oracle query function $g_y:\{0,1\}^n \to \{0,1\}$ satisfying

$$\begin{cases} g_y(j)=1, \text{if } a_j \leq y \\ g_y(j)=0, \text{if } a_j > y, \end{cases}$$

where $j \in \{0, 1, ..., N-1\}$. We then construct a unitary operation $U_{g_y}(\cdot)$ according to $g_y(\cdot)$. Using the operation $U_{g_y}(\cdot)$, instead of $U_f(\cdot)$, we can compute the number C, the number of j satisfying $g_y(j) = 1$. In the following, we list our algorithm to find the k-th smallest element in an unsorted database.

Step 1:

Set u=max and v=min.

Step 2:

We prepare two quantum registers. The first register with n qubits is initially set to $|0\rangle$. The second register with one qubit is also set to $|0\rangle$. We apply the Walsh-Hadamard Transformation H to $|0\rangle|0\rangle$. Then, we have

$$H(|0\rangle|0\rangle) \longrightarrow \frac{1}{\sqrt{2^n}} \sum_{j=0}^{2^n-1} |j\rangle|0\rangle.$$

Step 3:

Let $y = \left\lfloor \dfrac{u+v}{2} \right\rfloor$. We perform the operation $U_{g_y}(\cdot)$ which applies $g_y(\cdot)$ to the first register and then stores the result in the second register. Thus, we have

$$U_{g_y}(\frac{1}{\sqrt{2^n}} \sum_{j=0}^{2^n-1} |j\rangle|0\rangle) \to \frac{1}{\sqrt{2^n}} \sum_{j=0}^{2^n-1} |j\rangle|g_y(j)\rangle.$$

Step 4:

We measure the second register and compute the number C, the number of j satisfying $g_y(j) = 1$.

Step 5:

If C is less than k, then we set v=y; otherwise, we set u=y. If u=v+1, we have found the k-th smallest element u and output u; otherwise goto Step 2.

*Example:*

Given the unsorted database { $a_0 = 5$, $a_1 = 13$, $a_2 = 6$, $a_3 = 10$, $a_4 = 9$, $a_5 = 11$, $a_6 = 3$, $a_7 = 7$ }, we want to find the 4-th smallest element with the integer domain $D = [1..16]$. For simplicity, we assume that the measurement accuracy value $\varepsilon \approx 3$. Because $|D| = 16$, we list the following $\ln|D| = 4$ runs to discover the object element.

**Run 1:**

Step 1: Set u= 16 and v=1.

Step 2:

$$H(|0\rangle|0\rangle) \longrightarrow \frac{1}{2}\sum_{j=0}^{7}|j\rangle|0\rangle = \frac{1}{2}(|0\rangle|0\rangle + |1\rangle|0\rangle + |2\rangle|0\rangle + |3\rangle|0\rangle + |4\rangle|0\rangle + |5\rangle|0\rangle + |6\rangle|0\rangle + |7\rangle|0\rangle)$$

Step 3: Let $y = \left\lfloor \frac{16+1}{2} \right\rfloor = 8$.

We apply $U_{g_8}$ to the above state. We have

$$U_{g_8}(\frac{1}{2}\sum_{j=0}^{7}|j\rangle|0\rangle) = \frac{1}{2}(|0\rangle|1\rangle + |1\rangle|0\rangle + |2\rangle|1\rangle + |3\rangle|0\rangle + |4\rangle|0\rangle + |5\rangle|0\rangle + |6\rangle|1\rangle + |7\rangle|1\rangle).$$

Step 4:

We measure the second register and compute C=4.

Step 5:

Because $C \geq 4$, we set u=8. We goto Step 2 because $u \neq v+1$.

**Run 2:**

Step 2:

$$H(|0\rangle|0\rangle) \longrightarrow \frac{1}{2}\sum_{j=0}^{7}|j\rangle|0\rangle = \frac{1}{2}(|0\rangle|0\rangle+|1\rangle|0\rangle+|2\rangle|0\rangle+|3\rangle|0\rangle+|4\rangle|0\rangle+|5\rangle|0\rangle+|6\rangle|0\rangle+|7\rangle|0\rangle)$$

Step 3: Let $y = \left\lfloor \frac{8+1}{2} \right\rfloor = 4$.

We apply $U_{g_4}$ to the above state and obtain

$$U_{g_4}(\frac{1}{2}\sum_{j=0}^{7}|j\rangle|0\rangle) = \frac{1}{2}(|0\rangle|0\rangle+|1\rangle|0\rangle+|2\rangle|0\rangle+|3\rangle|0\rangle+|4\rangle|0\rangle+|5\rangle|0\rangle+|6\rangle|1\rangle+|7\rangle|0\rangle).$$

Step 4:

We measure the second register and compute C=1.

Step 5:

Because C<4, we set v=4. We goto Step 2 because $u \neq v+1$.

**Run 3:**

Step 2:

$$H(|0\rangle|0\rangle) \longrightarrow \frac{1}{2}\sum_{j=0}^{7}|j\rangle|0\rangle = \frac{1}{2}(|0\rangle|0\rangle+|1\rangle|0\rangle+|2\rangle|0\rangle+|3\rangle|0\rangle+|4\rangle|0\rangle+|5\rangle|0\rangle+|6\rangle|0\rangle+|7\rangle|0\rangle)$$

Step 3: Let $y = \left\lfloor \frac{8+4}{2} \right\rfloor = 6$.

We apply $U_{g_6}$ to the above state. We have

$$U_{g_6}(\frac{1}{2}\sum_{j=0}^{7}|j\rangle|0\rangle) = \frac{1}{2}(|0\rangle|1\rangle+|1\rangle|0\rangle+|2\rangle|1\rangle+|3\rangle|0\rangle+|4\rangle|0\rangle+|5\rangle|0\rangle+|6\rangle|1\rangle+|7\rangle|0\rangle).$$

Step 4:

We measure the second register and compute C=3.

Step 5:

Because C<4, we set v=6. We goto Step 2 because $u \neq v+1$.

**Run 4:**

Step 2:

$$H(|0\rangle|0\rangle) \longrightarrow \frac{1}{2}\sum_{j=0}^{7}|j\rangle|0\rangle = \frac{1}{2}(|0\rangle|0\rangle + |1\rangle|0\rangle + |2\rangle|0\rangle + |3\rangle|0\rangle + |4\rangle|0\rangle + |5\rangle|0\rangle + |6\rangle|0\rangle + |7\rangle|0\rangle)$$

Step 3: Let $y = \left\lfloor \frac{8+6}{2} \right\rfloor = 7$.

We apply $U_{g_7}$ to the above state. We have

$$U_{g_7}(\frac{1}{2}\sum_{j=0}^{7}|j\rangle|0\rangle) = \frac{1}{2}(|0\rangle|1\rangle + |1\rangle|0\rangle + |2\rangle|1\rangle + |3\rangle|0\rangle + |4\rangle|0\rangle + |5\rangle|0\rangle + |6\rangle|1\rangle + |7\rangle|1\rangle).$$

Step 4:

We measure the second register and compute C=4.

Step 5:

Because $C \leq 4$, we set u=7. We stop our algorithm because u=7=v+1.

After Run 4, our algorithm output the 4-th element 7.

## 4. Discussion

In our algorithm, many assumptions are given. Here, we discuss the possible solutions if these assumptions have been destroyed. First, we discuss that D is not a finite set of integers. Second, we discuss that the domain is not given before searching. Third, we must face the accurate problem when the measurement accuracy $\varepsilon$ is less than n. Fourth, we must modify the input size when N is not the power of 2. In addition to the above four cases, we give the reason why oracle query function $g_y(\cdot)$ can be implemented by a unitary operation and discuss the relation between D and N.

*Case 1:*

In general condition, the domain D may be a set of real numbers or integers. If the domain D belongs to integers, then our algorithm can correctly find the k-th smallest element. However, if the domain D belongs to real numbers, our algorithm finds the approximate value by setting $y = \frac{u+v}{2}$ in Step 3 of our algorithm. That is, the found value cannot be guaranteed to be the object element. We can perform more oracle queries to raise the accuracy of the approximate value. For example, we assume that the domain D contains real numbers between zero and one. If we want to search $\frac{1}{7} = 0.142857142857142857...$, we perform five oracle queries and get the output 0.15625. We then perform sixth query to obtain the output 0.140625.

*Case 2:*

In our algorithm, we assume that the domain is public information in the unsorted database. If the domain is unknown, we apply the concept of median-of three partitioning [18] to estimate it. We randomly choose two elements, called $u'$ and $v'$, in the unsorted database. We then perform Step 2 to Step 4 with $y=u'$ and $y=v'$, respectively. We assume that the output are $Cu'$ and $Cv'$, where $Cu' \geq Cv'$. If $Cv' \leq k \leq Cu'$, then the domain is from $v'$ to $u'$. If $k < Cv' \leq Cu'$, we select another $v'$ to satisfy $Cv' \leq k$. If $Cv' \leq Cu' < k$, we select another $u'$ to satisfy $Cu' \geq k$. Thus, our algorithm can find the k-th smallest element in the unsorted database with unknown domain.

*Case 3:*

In bulk quantum Turing machine [12], the measurement exists a measurement accuracy $\varepsilon$. This result causes the inaccuracy in each measurement. According to [1, PHV03], however, the accuracy can be enhanced by repeating our algorithm a number of times. The accuracy level scales with the square-root of the number of experimental trials [1]. Assume that $N_\delta$ is the number of experimental trials. If we want to obtain accurate measurement, $N_\delta$ must satisfy $\frac{1}{2^k} \times \frac{1}{\sqrt{N_\delta}} < \frac{1}{2^n}$. Thus, we have $N_\delta > 2^{2(n-k)}$. We require $O(2^{2(n-k)} \ln|D|)$ oracle queries to obtain the correct element. When $k \approx n$, we only require $O(\ln|D|)$ oracle queries. However, when $k \ll n$, we require $O(N^2 \ln|D|)$ oracle queries.

*Case 4:*

If N is not the power of 2, say $2^{n-1} < N < 2^n$, we add extra $2^n - N$ elements with the maximum value. We then prepare input states with n qubits in first register. Thus, our algorithm can determine the k-th element of original N elements.

*Case 5:*

Brüschweiler [1] show that the oracle query function $f(\cdot)$ can be implemented by the unitary operation if $f(\cdot)$ is described as a permutation function. The used oracle query function $g_y(\cdot)$ in our algorithm is also described as a permutation function with (n+1) qubits, including the first and second registers. In the following, we give an example to understand that the oracle query function $g_y(\cdot)$ can be described as a permutation function. Assume that we want to search the 2-th element in the four unsorted elements { $a_0$, $a_1$, $a_2$, $a_3$ } with the domain D. We assume that only $a_2$ and $a_3$ are less than y. We need two qubits, say $I_1$ and $I_2$, in the first register and one ancillary qubit, say $I_0$, in the second register. Then, we must construct an oracle query function $g_y(\cdot)$ satisfying $g_y(0)=0$, $g_y(1)=0$, $g_y(2)=1$, and $g_y(3)=1$. Thus, Fig. 1 show that $g_y(\cdot)$ operates on $I_1$ and $I_2$ with the output stored on $I_0$. Obviously, the action of $g_y(\cdot)$ can be viewed as a permutation of all states spanned by $I_1$, $I_2$, and $I_0$. Therefore, we can implement the unitary operation $U_{g_y}(\cdot)$ corresponding to the oracle query function $g_y(\cdot)$.

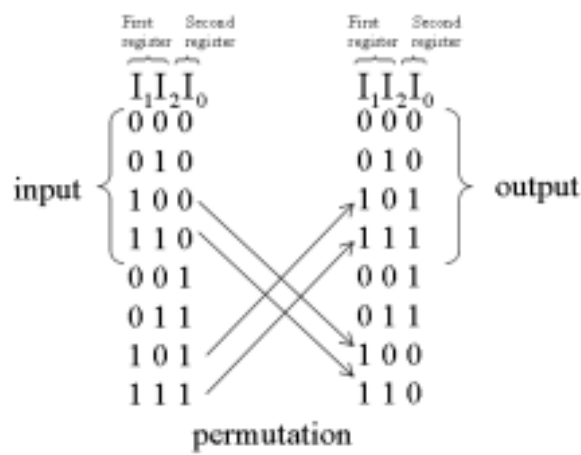

**Fig. 1:** Graphical representation of an oracle query function $g_y(\cdot)$ operating on $I_1$ and $I_2$ as a permutation using an ancillary bit $I_0$ with the output stored on $I_0$.

*Case 6:*

In this case, we discuss the relation of $|D|$ and N. When $|D| \approx N$, our algorithm requires $O(\ln N)$ oracle queries to find the target element. If the domain in a database is fixed and cannot vary with N, as the search of student's grade, our algorithm only requires constant oracle queries.

## 5. Conclusion

In this paper, we design our ensemble searching algorithm for the selection problem to find the k-th smallest element in the unsorted database. Given an unsorted database with $N=2^n$ elements, say $a_0, a_1, ..., a_{N-1}$, with the integer domain D, we construct oracle query functions $g_y(\cdot)$ such that $g_y(j)=1$ if $a_j \leq y$ and $g_y(j)=0$ if $a_j > y$, where y is the divided value in the domain D. This oracle function $g_y(\cdot)$ can be implemented by the corresponding unitary operation $U_{g_y}$ because $g_y(\cdot)$ can be described as a permutation function. We then use the ensemble counting scheme with the oracle query function $g_y(\cdot)$ to compute the number C, the number of j satisfying $g_y(j)=1$ in the unsorted database. According to C, we use "divide-and-conquer" strategy to adapt y until the object element is found. Our algorithm requires $O(2^{2(n-\varepsilon)} \ln|D|)$ oracle queries, where $\varepsilon$ is measure accuracy, to find the k-th element. When $\varepsilon \approx n$, our algorithm only requires $O(\ln|D|)$ oracle queries. Our algorithm also solves the media problem, the maximum problem and the minimum problem by setting k because these problems are special cases of the selection problem.